\documentclass[aps,reprint,groupedaddress,floatfix]{revtex4-2}

\usepackage{amsmath,amssymb}
\usepackage{mathrsfs}
\usepackage{graphicx}
\usepackage{array}

\begin{document}

\title{Algebraic models of plane Couette equilibria}

\author{Pratik P. Aghor}
\email[]{paghor1@jh.edu}
\affiliation{Department of Mechanical Engineering, Johns Hopkins University, Baltimore, Maryland, U.S.A}

\author{John F.\ Gibson}
\email[]{john.gibson@unh.edu}
\affiliation{Integrated Applied Mathematics Program, Department of Mathematics \& Statistics, University of New Hampshire, Durham, New Hampshire, U.S.A}

\date{\today}

\begin{abstract}
  Recent computations of weakly unstable equilibria, traveling waves, and periodic orbits in
  transitional shear flows suggest a spatiotemporal, dynamical-systems approach to low-Reynolds
  turbulence.
  Many invariant solutions have been computed precisely using high-dimensional direct numerical
  simulations, but little is known about how many solutions exist, how they are organized, or
  which sets of solutions best characterize the flow.
  In this paper we present a framework for addressing these questions in a low-dimensional
  context.
  Using classical approximation methods and exploiting symmetries and kinematic constraints,
  we derive ordinary differential equation models of plane Couette flow
  whose equilibria are governed by systems of quadratic algebraic equations. 
  Solutions of these algebraic systems approximate known equilibria of plane Couette flow 
  in as few as 17 dimensions and converge toward the known solutions as dimension
  increases.
  Searches over the systems produce sixteen distinct equilibrium solution branches in seven
  different symmetry groups. 
  These results suggest that the equilibrium and traveling-wave solutions of closed shear
  flows are organized by the algebraic structure of systems of quadratic equations.
  Additionally, the differential equations and divergence-free basis provide explicit,
  closed-form, and convergent dynamical-systems representations of plane Couette flow.

\end{abstract}

%\keywords{}

\maketitle

\newcommand{\bPsi}{{\bf \Psi}}
\newcommand{\be}{{\bf e}}
\newcommand{\bff}{{\bf f}}
\newcommand{\bg}{{\bf g}}
\newcommand{\bu}{{\bf u}}
\newcommand{\bx}{{\bf x}}
\newcommand{\ex}{\be_x}
\newcommand{\ey}{\be_y}
\newcommand{\ez}{\be_z}
\newcommand{\grad}{{\bf \nabla}}
\newcommand{\lapl}{\nabla^2}
\newcommand{\sleq}{\mkern-4mu \leq \mkern-4mu}

\newcommand{\mx}{x}
\newcommand{\bmx}{{\bf \mx}}
\newcommand{\sxy}{\sigma_{xy}}
\newcommand{\sz}{\sigma_{z}}
\newcommand{\sxyz}{\sigma_{xyz}}
\newcommand{\tx}{\tau_x}
\newcommand{\tz}{\tau_z}
\newcommand{\txz}{\tau_{xz}}
\newcommand{\Gpcf}{G_{\text{PCF}}}

\section{Introduction \label{sec:introduction}}

Dynamical systems theory provides a promising theoretical framework for the
development of dynamic, spatiotemporal models of turbulent flows
\citep{Hopf1942, Ruelle1971, LandfordARFM82, Holmes2012, ChaosBook}.
Ordinary differential equation (ODE) models have been developed for a variety
of transitionally turbulent flows, including Rayleigh-B\'enard convection
\citep{Lorenz1963}, the turbulent boundary layer \citep{Aubry1988, Holmes2012},
and plane Couette flow \citep{Waleffe1997,Moehlis2004,Smith2005,linot2023dynamics}.
Generally, in low dimensions ($O(10)$), these models exhibit qualitative similarity
to the underlying flows and provide insight to their behavior, while in higher dimensions
($O(100)$ and above), they begin to replicate some quantitative measures
\cite{Gibson2002, Smith2005}.
The neural-ODE models of plane Couette flow in \cite{linot2023dynamics} reproduce
short-term dynamics and global statistics with only $O(20)$ dimensions, at the cost
of a complex and inherently computational data-trained modeling framework.

A related line of research forgoes forgoes low-d modeling and treats direct numerical
simulations (DNS) of closed shear flows as very-high-dimensional dynamical systems, 
in order to compute to their equilibria, traveling waves, and periodic orbits directly
\citep{Nagata1990, Clever1992, Schmiegel1999, Kawahara2001, Wedin2004, Viswanath2007, 
Pringle2007, Gibson2009, Park2015, sharma2016low, AhmedPRE2020basis}.
That such invariant solutions exist, are computable, and have weak and relatively few
instabilities suggests that closed, transitional flows might be modeled as a pseudo-random 
walk within the network of low-dimensional unstable manifolds of a set unstable 
solutions \citep{Gibson2008,Cvitanovic2010}.
Many invariant solutions have been found, but little is known about their number or 
organization. Extreme dimensionality prevents comprehensive exploration of state
space and determination of sufficient sets of solutions to represent the global dynamics. 
%Methods for obtaining good initial guesses for solutions form an area of active research. 

Waleffe's theory of a self-sustaining process in plane Couette flow \citep{Waleffe1997}
stands at an interesting midpoint between these low- and high-dimensional approaches.
Waleffe showed that three spatial modes (roll, streak, and streak instability)
summed at proper amplitudes achieve an approximate force balance of the Navier-Stokes
equations. The balance is close enough to converge to a precise, high-dimensional DNS
equilibrium under numerical continuation \citep{Waleffe1997, Waleffe1998, Waleffe2001},
suggesting that the equilibrium is a low-dimensional force balance with perturbative
corrections in higher-order modes. 

In this paper we generalize Waleffe's self-sustaining process
by deriving algebraic force-balance equations from Galerkin projection of
Navier-Stokes onto sets of spatial modes that respect the flow's symmetries
and kinematic constraints. 
Our methods follow the ODE model derivations of \citep{Aubry1988, Holmes2012, Smith2005}
with a few key differences.
We use a divergence-free, periodic/no-slip basis set formed from elementary functions,
%xsimilar to that of \cite{Moser1983}, 
so that the basis and models can be expressed in closed form and extended to arbitrarily
high spatial resolution and model dimension.
The basis elements respect the symmetries of the flow, so that distinct ODE models can be
formed for the flow's distinct invariant symmetric subspaces, yielding relatively high spatial
resolution for relatively low dimension.
Lastly, we focus on the accuracy of the ODE equilibria compared to DNS, 
instead of the global dynamical behavior of the ODE models.

Our results include rough reproduction of the Nagata \citep{Nagata1990, Clever1992, Waleffe1998}
and Itano \& Genelaris \citep{Itano2009,Gibson2009} equilibria in $O(20)$ dimensions,
with convergence toward DNS solutions as model dimension increases.
Exhaustive searches of the models' state spaces produce new equilibrium solution branches
in each of seven symmetry groups that support equilibria. 
Additionally, the ODE models derived here provide a family of explicit, closed-form,
dynamical-systems models of plane Couette flow, parameterized by geometry, Reynolds number,
symmetry, and spatial resolution/dimensionality, which should prove useful for further
study of dynamical-systems representations of transitional flows.

\section{Methods \label{sec:methods}}

Following \cite{Gibson2008}, we express the nondimensionalized Navier-Stokes equations
for plane Couette flow as 
\begin{align}
  \frac{\partial \bu}{\partial t} + v \, \ex + y \frac{\partial \bu}{\partial x} + \bu \cdot \grad \bu &= -\grad p + \frac{1}{Re} \lapl \bu \label{eqn:NSE},\\
  \grad \cdot \bu &= 0. \nonumber
\end{align}
These equations result from expressing the total velocity field as the sum of the laminar base
flow and a fluctuating velocity, $\bu_{\text{tot}}(\bx,t) = y \, \ex + \bu(\bx,t)$, and substituting
this sum into the Navier-Stokes equations for $\bu_{\text{tot}}$. The streamwise, wall normal, and
spanwise spatial coordinates are $x,y,z$, and the components of the fluctuating velocity $\bu$ are
$u,v,w$. 
The computational domain is $\Omega = [0, L_x) \times [-1, 1] \times [0, L_z)$,
with periodic boundary conditions in $x$ and $z$, $\bu = 0$ at the walls $y = \pm 1$,
and mean pressure gradient fixed at zero.
%constraint on the mean flow, we fix the  %$\int_{\Omega} \grad p(\bx,t) \, d\Omega = 0$.
The laminar solution is then $\bu(\bx,t) = 0$.
The inner product and $L_2$ norm are $(\bff, \bg) = \frac{1}{|\Omega|}\int_{\Omega} \bff \cdot \bg \, d\Omega$
and $\| \bff \| = (\bff, \bff)^{1/2}$. 

Equations (\ref{eqn:NSE}) and their boundary conditions are equivariant under
the symmetries
\begin{align}
  \sxy &: [u,v,w](x,y,z) \rightarrow [-u,-v,w](-x,-y,z),\nonumber \\
  \sz &: [u,v,w](x,y,z) \rightarrow [u,v,-w](x,y,-z), \\
  \tau(a,b) &: [u,v,w](x,y,z) \rightarrow [u,v,w](x-a,y,z-b). \nonumber
\end{align}              
Subgroups generated from these symmetries
determine the invariant symmetric subspaces of the flow. The subgroups generated by
$\sxy, \sz,$ and the half-domain phase shifts $\tx = \tau(L_x/2,0)$ and $\tz = \tau(0, L_z/2)$
are enumerated and classified in \cite{aghor2025symmetry}. For notational compactness, we let 
$\sigma_{xyz} = \sigma_{xy} \sigma_z$ and $\tau_{xz} = \tau_x \tau_z$. 

We discretize (\ref{eqn:NSE}) by Galerkin projection onto a real-valued basis set
$\{\bPsi_{ijkl}(\bx)\}$ for velocity fields $\bu$ over $\Omega$ whose linearly independent
elements individually respect the boundary conditions and zero-divergence constraint.
The basis elements are formed from tensor products of real-valued $x,z$ Fourier modes
and polynomials in $y$. The construction is similar to that of \citep{Moser1983}, but
with basis elements that are (1) real-valued, (2) either symmetric or antisymmetric in
each of $\sxy, \sz, \tx,$ and $\tz$, and (3) explicitly defined and enumerated.
The $j,k$ indices of $\bPsi_{ijkl}$ specify $x,z$ wavenumbers in the real-valued Fourier
modes, with $j<0$ signifying $\cos(\alpha jx)$, $j>0$ signifying $\sin(\alpha j x)$,
and $j=0$ unity, and similarly for $k$ and $z$. 
The $l$ index governs the order of the polynomials in $y$.
The $i \in \{1,\ldots, 6\}$ index specifies one of six functional forms for $\bPsi_{ijkl}$.
Further details are given in the appendix.
We set the spatial resolution of the models by 
restricting indices to finite ranges $-J \sleq j \sleq J,\, -K \sleq k \sleq K,\; 0 \sleq l \sleq L$;
this produces a finite basis set with $(2J+1)(2K+1)(2L+1) + 1$ elements. 
Basis sets can also be restricted by symmetry; i.e. by selecting subsets that satisfy
the symmetries of a given symmetry subgroup.
Each generator of the subgroup reduces the basis set by a factor close to 2.
For example, the $J,K,L = 1,2,5$ basis  with no symmetry restrictions has $106$ elements,
whereas its $\langle \sxyz, \txz\rangle$-symmetric subset has $27$ elements.
After specifying a finite basis in terms of $J,K,L$ and a symmetry subgroup, we
then re-express the basis with a linear index: $\{\bPsi_n : n = 1,2, \ldots, m\}$.

The Galerkin projection is performed by substituting the expansion
$\bu(\bx, t) = \sum_{j=1}^{m} \mx_j(t) \bPsi_j(\bx)$ into (\ref{eqn:NSE}) and taking the
inner product of the resulting equation against $\bPsi_i$.
The pressure term vanishes due to the incompressibility and boundary conditions of the basis
elements. The resulting system is an ODE model of 
(\ref{eqn:NSE}) for the given symmetric subspace and spatial resolution,
\begin{align}
B \frac{dx}{dt} = A x + N(x), \label{eqn:ODE}
\end{align}
where $A$ and $B$ are $m \times m$ matrices, $B_{ij} = (\bPsi_i, \bPsi_j)$ and
$A_{ij} = (-\Psi_{i,v} \ex - y \, \partial \bu /\partial x + (1/Re) \lapl \bPsi_i, \bPsi_j)$,
$N(x)$ is a quadratic nonlinear term
$\left[N(x)\right]_i = \sum_{j,k=1}^m N_{ijk} x_j x_k$,
$N_{ijk} = (\bPsi_i, \bPsi_j \cdot \grad \bPsi_k)$,
and $x \in \mathbb{R}^m$ is the state-space vector of time-varying expansion coefficients $x_j$.
We use $x$ for both the state-space vector and the streamwise coordinate of the spatial domain; 
the distinction should be clear by context.
Each of $A$, $B$, and $N$ is sparse due to the orthogonality of the $x,z$ Fourier modes.
%$B$ is nonsingular due to the linear independence of the basis elements.
If we choose rational
numbers for $\alpha = 2\pi/L_x, \gamma=2\pi/L_z$, and $Re$, the elements of $A$,
$B$, and $N$ are also rational and can be computed exactly. The expansion coefficients $x$ of
a given velocity field $\bu$ are computed by solving the system $Bx = b$ where $b_i = (\bPsi_i, \bu)$.
This defines a projection operation on velocity fields, $P_{\bPsi}(\bu) = \sum_{j=1}^m x_j \bPsi_j$.
The Julia package CloudAtlas.jl \citep{cloudatlas} provides a mixed symbolic-numeric system for
computing the ODE models and their equilibria.

\section{Results \label{sec:results}}

\newcommand{\ODEeqb}{\bu^*_{\text{ODE}}}
\newcommand{\errp}[2]{#1 + #2}
\newcommand{\DNSeqb}{{\bf u}_{\text{DNS}}^*}
\newcommand{\myadd}[2]{{#1} + {#2}}

\begin{table}
\begin{tabular}{rrrr|lllc}
  J & K & L & m & $P_{\Psi}$ err &  EQ err & ~~$\lambda_1$ & bifurcation pt \\ 
\hline
  1 & 1 & 3 & 17 & ~$0.13 $ & $0.40$ & $0.07\pm0.13i$ & $(173.24, 2.768)$ \\
  1 & 2 & 3 & 27 & ~$0.079 $ & $0.24$ & $0.0588$ & $(175.63, 1.743)$\\
  1 & 3 & 5 & 59 & ~$0.037$ & $0.16$ & $0.0619$ & $(153.76, 1.628)$ \\
  %2 & 3 & 5 & 97 & ~$0.024$ & $0.12$       & $0.0573$ & $(158.42, 1.676)$ \\
  2 & 4 & 7 & 169 & ~$0.011$ & $0.036$ & $0.0510$ & $(158.79, 1.850)$ \\
  3 & 5 & 9 & 367 & ~$0.0043$ & $0.0058$ & $0.0509$ & $(163.04, 1.837)$\\
  3 & 6 & 11 & 524 & ~$0.0026$ & $0.0032$ & $0.0499$ & $(163.31, 1.831)$ \\
  4 & 7 & 13 & 912 & ~$0.0009$ & $0.0010$ & $0.0501$ & $(163.44, 1.835)$ \\
  \hline
  \multicolumn{6}{l}{Direct numerical simulation} & $0.0501$ &  $(163.52, 1.843)$\\
  %4 & 6 & 11 & 673 & ~$0.0016$ & $0.0022$ & 0.0500 &\\
  %5 & 7 & 13 & 1115 & ~6.4e-04 & 7.5e-04
\end{tabular}
\caption{Convergence of ODE equilibria to direct numerical simulation as a function of
  discretization parameters $J,K,L$ and ODE dimension $m$, for the Nagata lower-branch
  equilibrium. 
  The projection error $P_{\Psi} ~\text{err} = \|\DNSeqb - P_{\Psi} \DNSeqb \|/\|\DNSeqb\|$,
  equilibrium error, $EQ ~\text{err} = \|\DNSeqb - \ODEeqb \|/\|\DNSeqb\|$, and leading
  eigenvalue $\lambda_1$ are given for $Re=200$ and $L_x,L_z = 2\pi,\pi$.
  The $(Re,I)$ column shows the bifurcation point of each ODE solution.
  The last row shows the leading eigenvalue and bifurcation point computed with DNS, using 
  a $48 \times 49 \times 48$ discretization and 2/3-style dealiasing.
  \label{tbl:convergence}
}
\end{table}

First we compare equilibria of the ODE models to those of DNS
% for different discretization levels, symmetry groups, and Reynolds numbers.
by replicating the Nagata equilibrium with symmetry subgroup 
$\langle \sxyz, \sz\txz\rangle$. We chose a reference solution $\bu_{\text{DNS}}^*$
at $Re=200$, $L_x,L_z=2\pi, \pi$ computed with 
standard Fourier-Chebyshev spatial discretization and nonlinear solution algorithms
\cite{Canuto2007}. 
For ODE models, we specify discretization parameters $J,K,L$, construct the finite basis
$\{\bPsi_j : j = 1,\ldots,m\}$ for the $\langle \sxyz, \sz\txz\rangle$ subspace,
and compute $A,B$ and $N$ as described in \S \ref{sec:methods}.
Equilibria of the ODE model are solutions of $f(x) = B^{-1}(Ax + N(x)) = 0$,
a system of $m$ quadratic algebraic equations in $m$ unknowns. We find 
solutions $x^*$ using a trust-region Newton method and explicit
numerical representations of $f(x)$ and its derivative $D\!f(x)$.
%$[Df(x)]_{ij} = \partial f_i/\partial x_j = A_{ij} + \sum_{j,k=1}^m (N_{ijk} + N_{ikj}) x_k$.
Initial guesses for ODE equilibria are obtained by projection of the reference solution
onto the basis. The velocity field for an ODE equilibrium is $\ODEeqb = \sum_{j=1}^m x_j^* \bPsi_j$.

Table \ref{tbl:convergence} summarizes the results of this procedure for
a variety of discretization parameters $J,K,L$ and model dimensions $m$.
We found that choosing $L \approx 2K \approx 4J$ 
roughly optimized the projection error $\|\DNSeqb - P_{\Psi} \DNSeqb \|/\|\DNSeqb\|$
for a given dimension. 
The 17-dimensional 1,1,3 model was the lowest-dimensional model with an equilibrium near
the projection of the reference Nagata solution, with a projection error of about 13\%
and an equilibrium solution error $\|\DNSeqb - \ODEeqb \|/\|\DNSeqb\|$
of about 40\%. As $J,K,L$ and $m$ increase, both errors steadily decrease, reaching $10^{-3}$ accuracy for 
$m \approx 10^3$.

\begin{figure}
  \includegraphics[width=0.45\textwidth]{"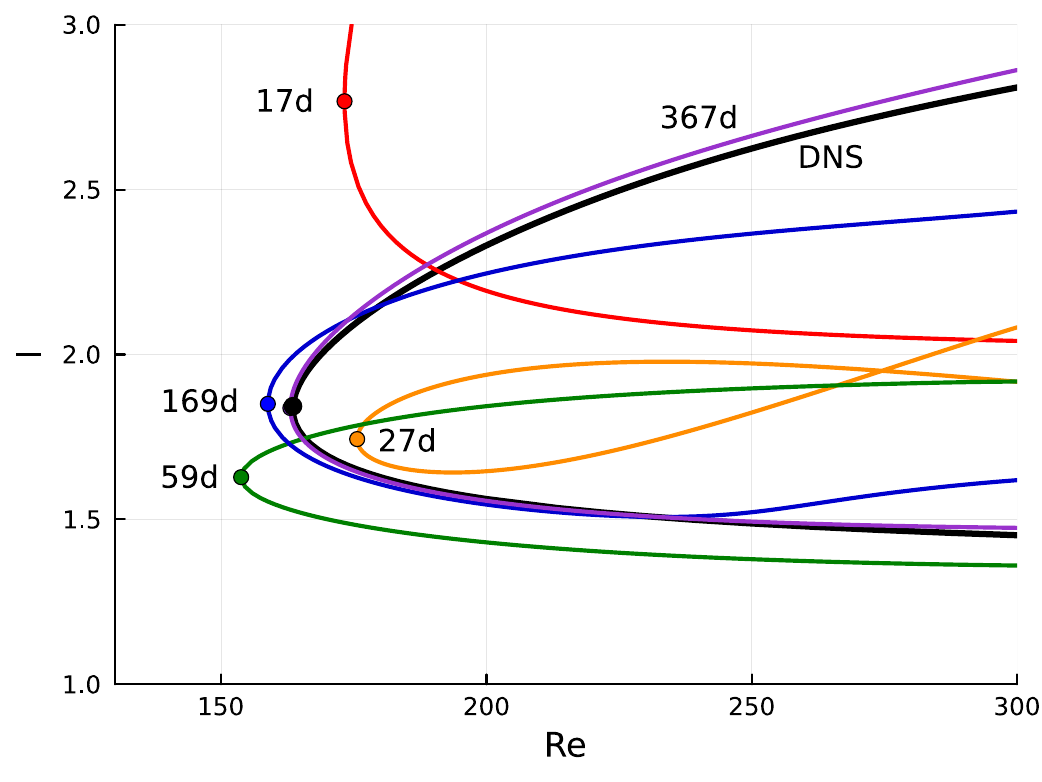"}
  \caption{Bifurcation diagram of ODE equilibria compared to DNS for the Nagata 
   equilibrium at $L_x,L_z=2\pi,\pi$, wall shear rate $I$ versus Reynolds
   number $Re$. ODE solution curves are labeled with their dimensionality;
   e.g. 17d for the $J,K,L,m = 1,1,3,17$ ODE solution. 
   Bifurcation points of the solutions are marked with circles; their $(Re,I)$
   locations are listed in Table \ref{tbl:convergence}. 
   \label{fig:bifurcation}
  }
\end{figure}

Figure \ref{fig:bifurcation} shows the bifurcation curves of the ODE equilibria
compared to DNS, computed with BifurcationKit.jl \citep{veltz:hal-02902346} and
channelflow-2.0 \cite{channelflow}. 
The ODE solutions appear in saddle-node bifurcations in the
range $153 < Re < 174$, compared to the critical Reynolds number $Re=163.52$ of the DNS.
The ODE bifurcation curves converge on the DNS curve as dimension increases, with the
367-dimensional $J,K,L = 3,5,9$ ODE model accurate to three digits in the bifurcation
point and tracking the DNS curve to a few percent up to $Re=300$.
\begin{figure}
  (a) \includegraphics[width=0.2\textwidth]{"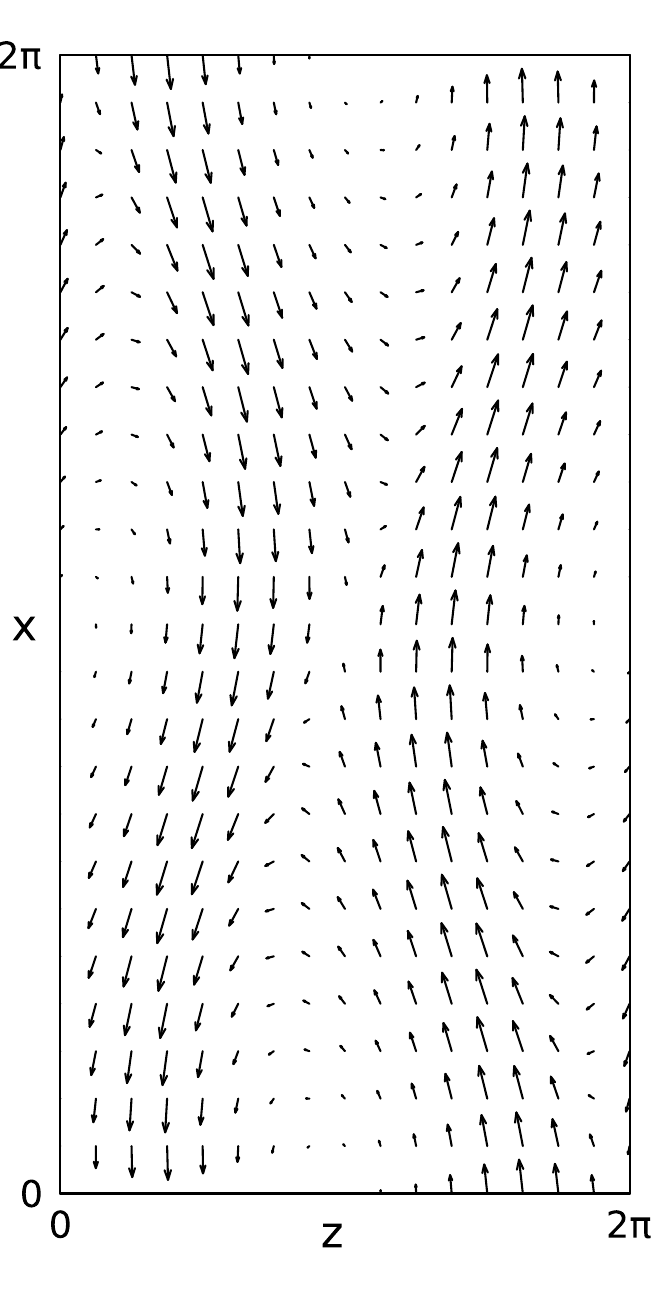"} 
  (b) \includegraphics[width=0.2\textwidth]{"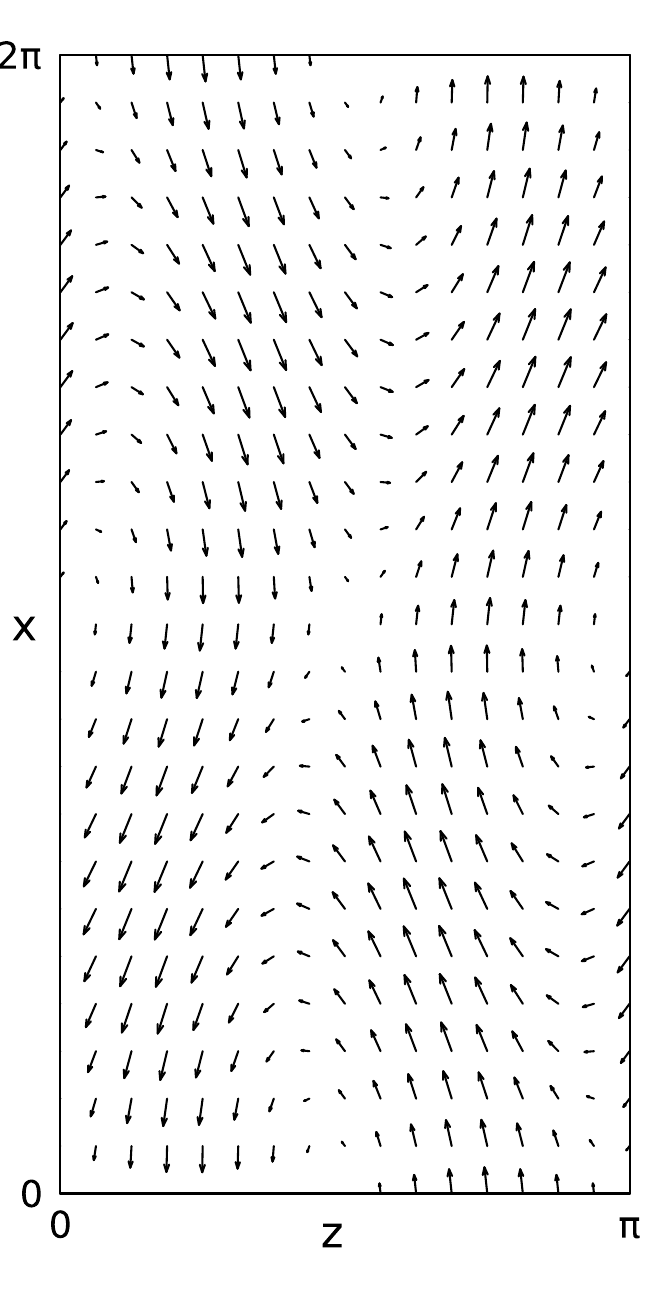"} \\
  (c) \hspace{-2mm} \includegraphics[width=0.215\textwidth]{"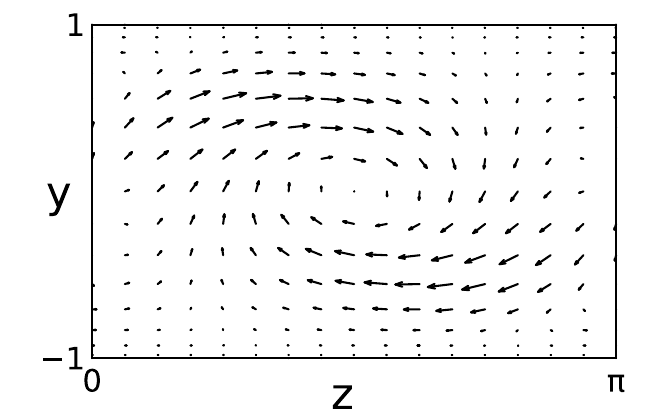"}
  \hspace{-2mm} (d) \hspace{-2mm} \includegraphics[width=0.215\textwidth]{"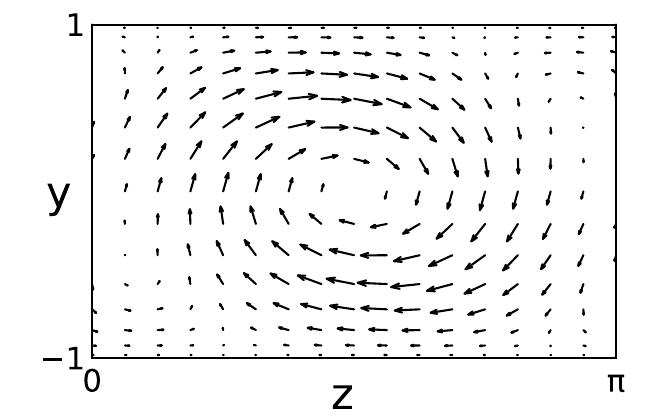"}
  \caption{Velocity fields of the Nagata equilibrium, (a,c) DNS versus
    (b,d) 17d ODE model. Figures (a,b) show $u,w$ velocity in the $y=0$ midplane; (c,d) show
    $v,w$ velocity at $x=0$. The solutions are shown at their bifurcation points,
    $(Re, I) = (163.5, 1.84)$ for DNS and $(Re,I) = (173.2, 2.77)$ for the 17d ODE model.
  }
  \label{fig:velocity}
\end{figure}

Figure \ref{fig:velocity} compares the velocity field of the 17d ODE equilibrium and
the reference DNS solution at their respective bifurcation points. Though the 
relative difference between these fields is about 40\%, the ODE equilibrium clearly replicates
the wobbly roll-streak structure of the reference solution.
These results suggest that the algebraic system is a rough but faithful
low-dimensional model of the large-scale structure and pointwise force balance of the Nagata solution.
We obtained similar results for the Itano \& Generalis solution at the same flow
parameters.

% , while the higher-dimensional models verge toward direct numerical simulation.

\begin{figure}
  \includegraphics[width=0.45\textwidth]{"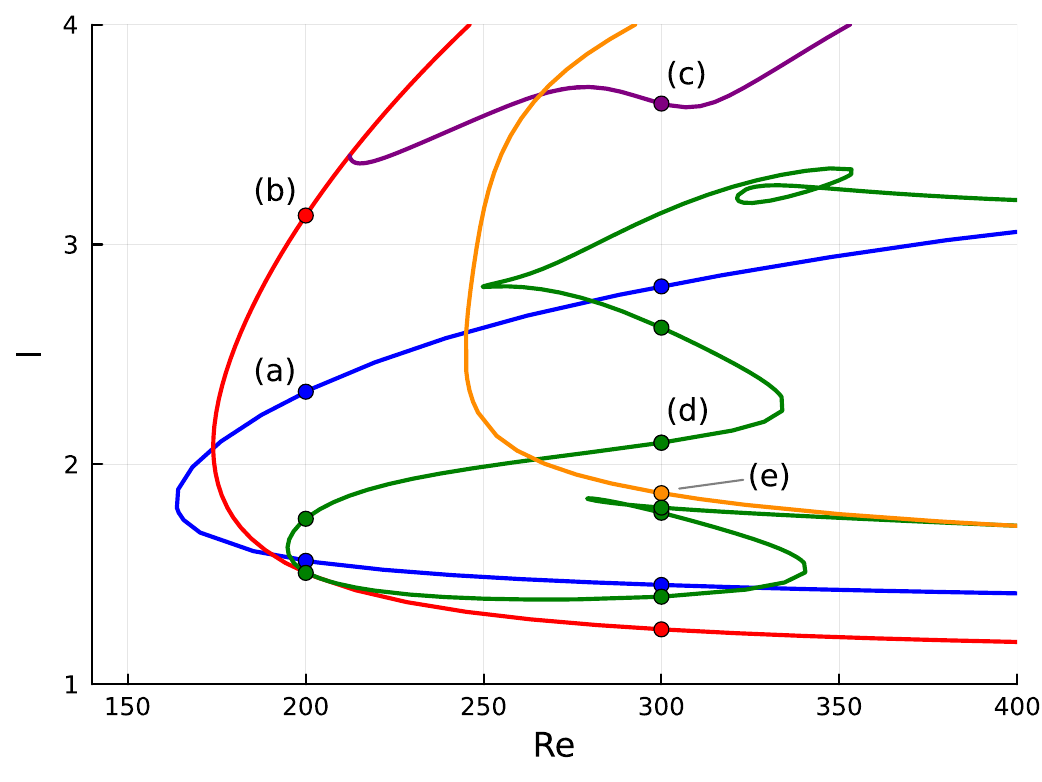"}
  \caption{Bifurcation diagram for plane Couette equilibria, wall shear rate $I$
    versus Reynolds number $Re$, computed with DNS. 
    Dots represent the DNS equilibria found from nearby solutions of 
    $\langle \sxy \tz, \sz \rangle$-symmetric ODE models at $L_x,L_z=2\pi,\pi$,
    $Re=200$ and $300$. 
    (a) The Nagata solution (EQ1),  (b) the Itano \& Generalis solution (EQ7), (c,d,e) new
    solution branches. Branches (a,e) have an additional $\txz$ symmetry.
}
   \label{fig:fuzzing}
\end{figure}

Next we search for equilibrium solutions of the ODEs and refine them with DNS. 
We searched for ODE equilibria at $L_x,L_z=2\pi,\pi$ and
$Re=200$ and $300$ in each of the seven half-shift symmetric subspaces that support equilibria
\citep{Gibson2009,aghor2025symmetry}, with randomly distributed initial guesses in
discretizations ranging from $(J,K,L) = (1,2,3)$ and $m\approx 25$ to $(3,6,11)$ and
$m \approx 500$. 
One simple and effective approach was to set each $x$ coefficient to a
uniform random number over $[-0.1,0.1]$ and then to rescale the modes that contribute to
the mean shear rate to fall randomly in $1 \leq I \leq 3$.
In a typical run, a few hundred of a thousand random initial guesses would converge
onto a handful of distinct ODE equilibria, with some to most these converging
when refined with DNS, depending on model resolution. 
Figure \ref{fig:fuzzing} shows the results of such searches of the
$\langle \sigma_{xyz}, \sigma_z \tau_{xz}\rangle$ subspace at $L_x, L_z$ and $Re=300$,
using discretizations from $(J,K,L,m) = (1,2,3,27)$ at $Re=200$ up to $(3,6,11,524)$
at $Re=300$. 
These searches found five distinct solution branches (a)--(e),
identifying five points on three branches at $Re=200$ and twelve points on five branches
at $Re=300$ (two with $I>4$ are not shown). Low-order ODE discretizations % such as $(J,K,L) = (1,2,3)$
easily reproduced the Nagata and Itano \& Generalis solutions, however with many spurious
ODE solutions that did not converge onto equilibria of DNS. Higher discretizations produced
fewer spurious ODE solutions and more accurate initial guesses for the DNS searches.
Following the same procedure in each of the seven symmetry groups at $Re=300$,
we found 25 points on 16 distinct solution branches, some replicating known solutions,
but most new. %These results will be presented in an expanded publication.
Computational costs for the ODE searches ranged from $O(10^{-3})$ cpu-seconds for
$m\approx 20$ to $O(1)$ cpu-seconds for $m \approx 500$, compared to $0(5000)$
cpu-seconds for each DNS search with $48 \times 49 \times 48$ Fourier-Chebyshev
discretization, 2/3-style deliasing, and $m \approx 10^5$ free variables. 
It is notable that the Nagata and Itano \& Generalis equilibria were the most frequently
found solutions in our searches and also required the fewest spatial modes. 
% and have the fewest unstable eigenvalues,
Whether these
solutions are special in some sense or if these differences are an artifact of the
particular parameters explored here is a matter for future research.

\section{Discussion \label{sec:discussion}}

We have shown that large-amplitude, nonlinear solutions of transitional plane Couette flow
are well-approximated by low- to moderate-dimensional quadratic algebraic equations, and
that these reduced equations form effective models for probing the state-space of the
fluid flow to find its equilibrium solutions. The quadratic systems are vastly simpler
than the Navier-Stokes equations and DNS in dimensionality, computational cost, and mathematical
tractibility.
The symmetries of the basis elements allow for the symmetric subspaces of the flow
to be explored independently with reduced dimensionality. 
Our numerical results show that several well-studied equilibria of plane Couette flow
are essentially low-dimensional force balances between the large-scale, low-order modes
allowed by geometry and kinematics, with perturbative corrections in the higher-order modes. 
As such the models form an algebraic, symmetry-based generalization of
Waleffe's self-sustaining process and an efficient method for exploring
the space of solutions as a function of parameters and symmetry group.
We expect that these methods can be extended to plane Poiseuille and pipe flows.
For planar shear flows, the divergence-free, no-slip, symmetric basis should be
useful in applications requiring projections, such as dynamic mode decomposition
\citep{linot2023dynamics}, state-space portraits \cite{Gibson2008}, and machine learning
\citep{Srinivasan2019}.

The methods presented here also provide a framework for representing transitional planar
shear flows with unconstrained ODEs in $\mathbb{R}^m$. Our 
results suggest that these ODEs are quantitatively accurate in relatively low dimensions
for the smooth portions of state space where low- to moderate-dissipation equilibria lie.
What dimensions are required to form faithful models globally is not yet clear. Studies of
time integrations of the ODEs and convergence of high-dissipation equilibria
and periodic orbits should clarify. Our preliminary results with time integrations
show qualitatively correct behavior in $O(100)$ dimensions --relatively high for dynamical-systems
theory, but substantially smaller than the dimensionality of DNS.
These unconstrained, and explicit ODE models lend themselves to analytic
and computational approaches that are practically impossible with DNS.
The coefficients in the matrices and nonlinear operators of ODE models provide explicit dynamical
linkages between the spatial modes of the velocity field. Statistics of the modes and linkages
during time evolution might suggest more efficient, data-driven truncations of the models than
the rectangular $J,K,L$ limitations used here, and lead to fine-grained and mode-flexible
reduced-order models.

The relatively low-dimensional quadratic equations governing the equilibria of the ODE
models suggest algebraic organization in the invariant solutions of the Navier-Stokes
equations for transitional shear flows. Between a small number of lowest-order
modes of a system, there can be only so many nonlinear force balances,
particularly when the nonlinear coupling is quadratic and sparse. The convergence
of the ODE equilibria to high-resolution simulations also suggests a route to understanding
invariant solutions of Navier-Stokes as limits of sequences of finite-dimensional
algebraic systems.

%\clearpage

\appendix*
\section{The divergence-free basis set \label{app:basis}}

%\begin{widetext}
\begin{table*}
%\centering
  \caption{Definition of basis elements $\bPsi_{ijkl}(\bx) = \Psi_{ijkl,u}(\bx) \, \be_x + \Psi_{ijkl,v} (\bx) \, \be_y + \Psi_{ijkl,w}(\bx) \, \be_z$.
    The $E_j$ and $E_k$ functions  represent real-valued Fourier modes, with $E_j(\alpha x) =  \cos(\alpha j x), 1,$
    and $\sin(\alpha j x)$ for  $j<0$, $j=0$, and $j>0$ respectively, and similarly for $E_k(\gamma z)$. The $S_{l}$
    functions are $(l+3)$th-order polynomials: $S_0(y) = y-y^3/3$ and $S_l(y) = (1-y^2)^2 P_{l-1}(y)$ for
    $l \geq 1$, where $P_l(y)$ is the $l$th Legendre polynomial.
  } %  \centering
\begin{tabular}{l|rrr|lll}
  & $\Psi_{ijkl,u}$ & $\Psi_{ijkl,v}$ & $\Psi_{ijkl,w}$ & \multicolumn{3}{c}{~index restrictions}\\ \hline
$\bPsi_{1jkl}$ & $E_k(\gamma z)\, S_l'(y)$ & 0~~ & 0~~ & $~~j \!=\! 0$ & & \\
$\bPsi_{2jkl}$ & 0~~ & $\gamma k \; E_{k}(\gamma z)\,  S_{l}(y)$ & $E_{-k}(\gamma z)\, S_{l}'(y)$ & $~~j\!=\!0,$ & $k\!\neq\!0,$ &$l\!\neq\!0$ \\
$\bPsi_{3jkl}$  & 0~~ & 0~~ & $E_j(\alpha x)\, S_l'(y)$ & & $k\!=\!0$ & \\
$\bPsi_{4jkl}$  & $E_{-j}(\alpha x)\, S_{l}'(y)$ & $\alpha j \; E_{j}(\alpha x)\,  S_{l}(y)$ & 0~~ &  $~~j\!\neq\!0,$ & $k\!=\!0,$ & $l\!\neq\!0$ \\
$\bPsi_{5jkl}$  & $\gamma k \; E_{-j}(\alpha x) E_k(\gamma z)\, S_l'(y)$ & 0~~ & $~~-\alpha j \; E_j(\alpha x) E_{-k}(\gamma z)\, S_l'(y)$ & $~~j \!\neq\! 0,$ & $k \!\neq\! 0$ \\
$\bPsi_{6jkl}$  & $\gamma k \; E_{-j}(\alpha x)\, E_k(\gamma z)\, S_{l}'(y)$ & ~~$2 \alpha \gamma j k \; E_{j}(\alpha x)\, E_k(\gamma z) \; S_{l}(y)$ & $\alpha j \; E_{j}(\alpha x)\, E_{-k}(\gamma z) S_{l}'(y)$ & $~~j\!\neq\!0,$ & $k\!\neq\!0,$ & $l\!\neq\! 0$
\end{tabular}
\label{tbl:basis_functions}
\end{table*}
%\end{widetext}

Table \ref{tbl:basis_functions} defines the basis functions 
$\bPsi_{ijkl}(\bx)$.
% = \Psi_{ijkl,u}(\bx) \, \be_x + \Psi_{ijkl,v}(\bx) \, \be_y + \Psi_{ijkl,w}(\bx) \, \be_z$.
The basis functions are designed (1) to individually satisfy incompressibility and
the boundary conditions, (2) to be real-valued, linearly independent, and complete, 
(3) to have the same tensor-product structure as typical Fourier-polynomial
representations in DNS but (4) without the complex
symmetries and linear dependencies that result from complex Fourier representations of
real-valued functions, and (5) to be individually symmetric or
antisymmetric with respect to each of $\sxy, \sz, \tx$, and $\tz$.

The $i$ index in $\bPsi_{ijkl}(\bx)$ governs which of six functional forms are needed
to span the space of divergence-free, no-slip velocity fields. 
Real-valued Fourier modes in $x$ and $z$ are represented by $E_j(\alpha x)$ and $E_k(\gamma z)$,
where $E_j(\alpha x) =  \cos(\alpha j x), 1,$ and $\sin(\alpha j x)$ for  $j<0$, $j=0$,
and $j>0$ respectively, and similarly for $E_k(\gamma z)$.
The differentiation rules are $E_j'(\alpha x) = \alpha j E_{-j}(\alpha x)$ and $E_k'(\gamma z) = \gamma k E_{-k}(\gamma z)$.
The $S_l(y)$ functions are odd/even polynomials in $y$, with definition $S_0(y) = y-y^3/3$ and
$S_l(y) = (1-y^2)^2 P_{l-1}(y)$ for $l \geq 1$, where $P_l(y)$ is the $l$th Legendre polynomial.
Thus, for example, $\bPsi_{2,0,3,1}(\bx) = 3 \gamma \sin(3 \gamma z) (1-y^2)^2 \be_y - \cos(3 \gamma z) 4y(1-y^2) \be_z$.

The stated goals for the basis set are achieved with these definitions.
In this discussion we drop $ijkl$ subscripts on basis elements $\bPsi$ for brevity.
The incompressibility of each basis element is clear from the differentiation
rules, and the $x,z$ periodicity from the definitions of $E_j(\alpha x)$ and $E_k(\gamma z)$. 
The no-slip boundary conditions at the walls are ensured by the definition of
$S_l(y)$. For the wall-normal velocity component $\Psi_v$, the wall boundary
conditions are $\Psi_v(x,\pm 1,z) = \partial /\partial y \, \Psi_v(x,\pm 1,z) = 0$. 
To represent the $y$ variation of $\Psi_v$, we use a polynomial basis
$\{S_l(y) : l = 1, 2, \ldots\}$. The $l$th element $S_l(y)$ is an $(l+3)$th-order
polynomial odd/even in $y$ with even/odd $l$. The factor of $(1-y^2)^2$ in the 
definition of $S_l(y)$ for $l>1$ ensures each element satisfies the four boundary
conditions $S_l(\pm 1) = S_l'(\pm) = 0$. The lowest-order element of the set is 
the 4th-order $S_1(y) = (1-y^2)^2$, a polynomial of the lowest order order that
can meet the four boundary conditions. The order of the polynomials in the set
increases by 1 for each successive element, ensuring that the set is linearly
independent and spans the space of polynomials that meet the four boundary conditions. 

The streamwise and spanwise components $\Psi_u$ and $\Psi_w$ each have two boundary
conditions at the walls, $\Psi_{u}(x,\pm 1,z) = 0$. The basis set
$\{S_l'(y) : l = 0, 1, \ldots\}$ represents the $y$ variation of these 
components. 
We have $S_0'(y) = 1-y^2$ and $S_l'(y) = 4y (1-y^2) P_{l-1}(y) + (1-y^2)^2 P_{l-1}'(y)$
for $l \geq 1$, so that $S_l'(y)$ is an $(l+2)$th-order even/odd polynomial in $y$
with even/odd $l$, and $S_l'(\pm 1) = 0$ for all $l$. 
The lowest-order element of this set is the 2nd-order $S'_0(y) = 1-y^2$,
a polynomial of the lowest order that can meet the two boundary conditions.
The increase of polynomial order by 1 in successive elements ensures that the
set is linearly independent and spans the space of polynomials that meet the
two boundary conditions.

%\begin{acknowledgments}
%\end{acknowledgments}

%\bibliographystyle{apsrev4-2}
\bibliography{algebraic-eqbs}

\end{document}